\documentclass[]{spie}  

\usepackage{amsmath,amsfonts,amssymb}
\usepackage[frozencache,cachedir=.]{minted}
\usepackage{graphicx}
\usepackage{listings}
\usepackage{float}
\usepackage{filecontents}
\usepackage{tabularx}
\usepackage{multirow}
\usepackage{multicol}
\usepackage{booktabs}
\usepackage{adjustbox}
\usepackage{cite} 
\usepackage{times}
\usepackage{epsfig}
\usepackage{amsmath}
\usepackage{amssymb}
\usepackage{mwe}
\usepackage{acro}
\usepackage{amssymb}
\usepackage{xcolor,colortbl}
\usepackage{relsize}
\usepackage{pifont}
\usepackage{float}
\usepackage{xcolor}
\usepackage[colorlinks=true, allcolors=blue]{hyperref}

\title{Eosinophils Instance Object Segmentation on Whole Slide Imaging Using Multi-label Circle Representation}

\author[a]{Yilin Liu}
\author[a]{Ruining Deng}
\author[a]{Juming Xiong}
\author[b]{Regina N Tyree}
\author[c]{Hernan Correa}
\author[b]{Girish Hiremath}
\author[d]{Yaohong Wang}
\author[a,d,e]{Yuankai Huo}

\affil[a]{Department of Computer Science, Vanderbilt University, Nashville, TN, USA}
\affil[b]{Division of Pediatric Gastroenterology, Hepatology, and Nutrition, Vanderbilt University Medical Center, Nashville, TN, USA}
\affil[c]{Department of Pediatrics, Vanderbilt University Medical Center, Nashville, TN, USA}
\affil[d]{Department of Pathology, Microbiology and Immunology, Vanderbilt University Medical Center, Nashville, TN, USA}
\affil[e]{Department of Electrical and Computer Engineering, Vanderbilt University, Nashville, TN, USA}

\authorinfo{Corresponding author: Yuankai Huo: E-mail: yuankai.huo@vanderbilt.edu}

\begin{document} 
\maketitle

\begin{abstract}
Eosinophilic esophagitis (EoE) is a chronic and relapsing disease characterized by esophageal inflammation. Symptoms of EoE include difficulty swallowing, food impaction, and chest pain which significantly impact the quality of life, resulting in nutritional impairments, social limitations, and psychological distress. The diagnosis of EoE is typically performed with a threshold (15 to 20) of eosinophils (Eos) per high-power field (HPF). Since the current counting process of Eos is a resource-intensive process for human pathologists, automatic methods are desired. Circle representation has been shown as a more precise, yet less complicated, representation for automatic instance cell segmentation such as CircleSnake approach. However, the CircleSnake was designed as a single-label model, which is not able to deal with multi-label scenarios. In this paper, we propose the multi-label CircleSnake model for instance segmentation on Eos. It extends the original CircleSnake model from a single-label design to a multi-label model, allowing segmentation of multiple object types. Experimental results illustrate the CircleSnake model's superiority over the traditional Mask R-CNN model and DeepSnake model in terms of average precision (AP) in identifying and segmenting eosinophils, thereby enabling enhanced characterization of EoE. This automated approach holds promise for streamlining the assessment process and improving diagnostic accuracy in EoE analysis. The source code has been made publicly available at \url{https://github.com/yilinliu610730/EoE}. 

\end{abstract}


\section{INTRODUCTION}
\label{sec:intro}  
Eosinophilic esophagitis (EoE) is a chronic and relapsing disease characterized by eosinophil-predominant inflammation in the esophagus. Clinically, patients with EoE often experience recurring symptoms such as difficulty swallowing, food impaction, chest pain, and heartburn~\cite{dellon2013acg}. These symptoms can severely impact an individual's quality of life, leading to impaired nutrition, social limitations, and psychological distress~\cite{papadopoulou2014management}. The diagnosis of EoE is confirmed by demonstrating at least 15 eosinophils (Eos) per high-power field (or approximately 60 Eos per mm2) in one of the multiple esophageal biopsies~\cite{liacouras2011eosinophilic,dellon2012diagnosis,muir2021eosinophilic}. 

A significant challenge in EoE diagnosis is the labor-intensive and time-consuming manual process of annotating and evaluating the presence and severity of the cellular and tissue alterations~\cite{czyzewski2021machine,adorno2021advancing,dellon2014phenotypic}. Additionally, gastroenterologists face difficulty in predicting disease progression, determining clinical phenotypes, and formulating personalized treatment plans for patients. Deep learning tools and models have been shown to improve the diagnostic accuracy, objectively analyzing clinical and histological data, and developing personalized treatment plans~\cite{saturi2023review}. However, this powerful approach has not been fully exploited to assist pathologists and gastroenterologists to advance the diagnosis and management of EoE patients~\cite{czyzewski2021machine}.

Many prior studies have investigated the deep learning based method for Eos quantification. Adorno III et al., conducted the first study to employ a U-Net model for Eos segmentation and EoE diagnosis~\cite{adorno2021advancing}. Subsequent studies focused on more fine grained clinical outcomes~\cite{vande2022utilizing,javaid2021deep,archila2023performance}, more advanced network architecture design~\cite{daniel2021pecnet,wu2020expert,wang2021eosinophil}, more efficient learning design~\cite{shi2022eosinophilic,pan2022object,alzu2021new}, and more detailed segmentation results~\cite{daniel2022deep,archila2022development,larey2022harnessing}. 
All aforementioned instance detection and segmentation methods utilized box representations, which could potentially result in less optimal performance compared to the recently introduced circle representation~\cite{yang2020circlenet,nguyen2021circle,nguyen2022circlesnake}.

\begin{figure}[ht]
  \centering
  \includegraphics[width=0.8\textwidth]{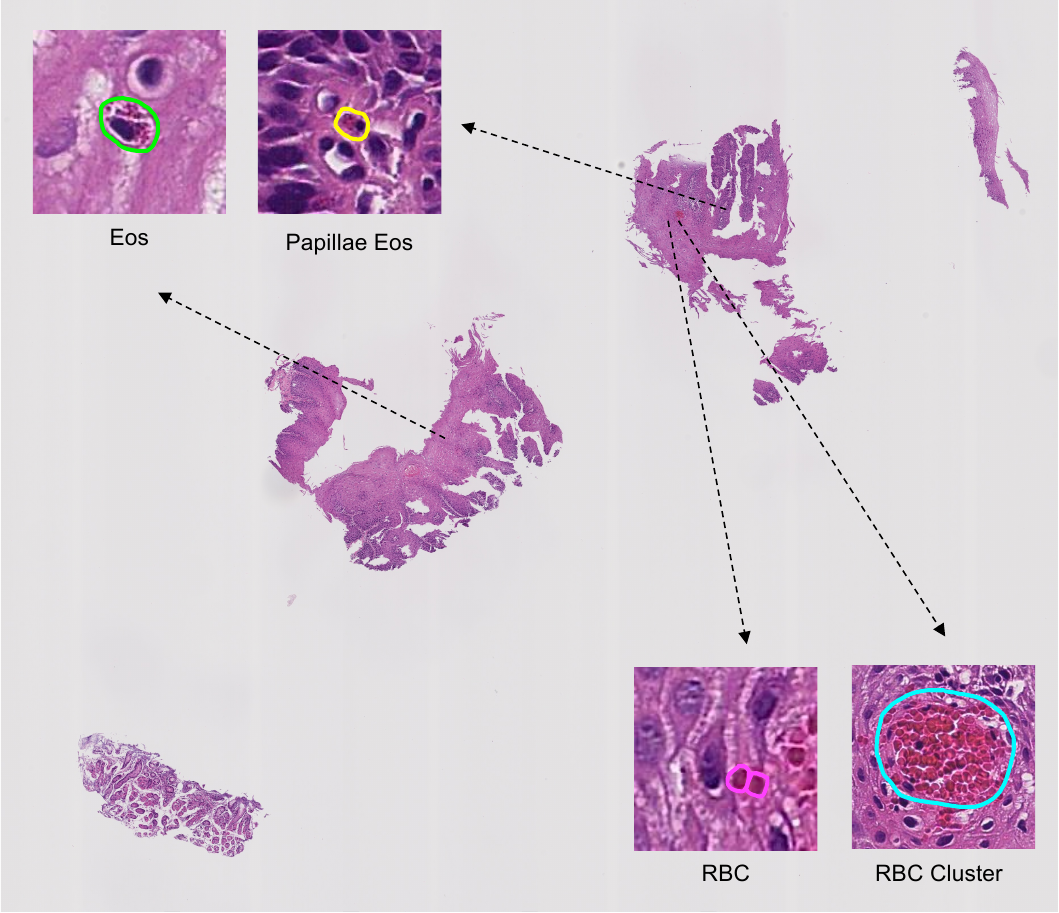}
  \caption{Four phenotypes for training a multi-label CircleSnake. The four phenotypes are Eos, Papillae Eos, RBC, and RBC Cluster. The last three phenotypes serve as ``hard negative" cases that can be confused with Eos.}
  \label{fig:phenotypes}
\end{figure}

In this paper, we present an advanced iteration of the CircleSnake model, elevating its functionality from single-label to multi-label instance segmentation. Our approach is tailored for the precise instance object segmentation of biomarkers in EoE and comprehensive analysis. It builds upon the CircleSnake model's foundation as a straightforward end-to-end circle contour deformation-based segmentation method for ball-shaped medical objects, seamlessly integrating deep learning tools with medical image analysis~\cite{nguyen2022circlesnake}. We demonstrate the effectiveness of the multi-label CircleSnake model in accurately identifying and quantifying the presence of Eos in entire esophageal biopsy Whole-Slide Images (WSIs). Compared to the traditional train-from-scratch Mask R-CNN model and the prevalent DeepSnake model counterpart, CircleSnake achieves superior average precision scores. Furthermore, our trained model successfully distinguishes four different types of cells associated with EoE, as depicted in Fig.\ref{fig:phenotypes}. This advancement of CircleSnake enables a more comprehensive and precise analysis of Eos in esophageal biopsies within the context of EoE.

To sum up, our contribution is four-fold: (1) We propose a new multi-label CircleSnake model that extends its capabilities from single-label instance segmentation to multi-label instance segmentation. (2) To the best of our knowledge, the proposed method is the first multi-label circle representation based instance object segmentation method.
(3) The proposed method yields superior average precision (AP) in identifying and segmenting Eos, compared to the prevalent Mask R-CNN and DeepSnake model in eosinophil segmentation. (4) The detailed process of "annotation to model" and the corresponding source code are now accessible to the public for reference.

\section{RELATED WORK}
\textbf{Advancing Eosinophilic Esophagitis Diagnosis and Phenotype Assessment with Deep Learning Computer Vision:} The paper utilizes a U-Net model for deep image segmentation to quantify Eos and diagnose EoE. It is the first study to employ a deep learning computer vision approach for EoE diagnosis, providing an automated process for disease severity tracking and assessment ~\cite{adorno2021advancing}.

\textbf{Utilizing Deep Learning to Analyze Whole Slide Images of Colonic Biopsies for Associations Between Eosinophil Density and Clinicopathologic Features in Active Ulcerative Colitis:} 
The paper utilizes a deep learning algorithm to identify Eos in colonic biopsies of ulcerative colitis patients with active disease and explores associations with histologic activity and clinical features. This automated approach provides valuable insights into disease pathogenesis and therapeutic response, aiding in disease severity assessment and treatment planning at diagnosis~\cite{vande2022utilizing}.

\textbf{Machine Learning Approach for Biopsy-Based Identification of Eosinophilic Esophagitis Reveals Importance of Global features:} This paper introduces a deep convolutional neural network (DCNN) approach combined with a systematic downscaling and cropping technique to accurately classify esophageal biopsies with EoE and reveals global histologic features. The use of AI-based computer vision analysis for EoE diagnosis offers a standardized and efficient method, with potential applications in other biopsy-based histologic diagnostics~\cite{czyzewski2021machine}.

\textbf{PECNet: A Deep Multi-Label Segmentation Network for Eosinophilic Esophagitis Biopsy Diagnostics: } The paper introduces PECNet, a deep multi-label segmentation network, for EoE biopsy diagnostics. PECNet successfully identifies and quantifies Eos in esophageal biopsies, providing an accurate and efficient automated method for EoE diagnosis and disease monitoring, addressing the challenges in digital pathology and small feature detection~\cite{daniel2021pecnet}.

\textbf{A Deep Multi-Label Segmentation Network For Eosinophilic Esophagitis Whole Slide Biopsy Diagnostics: } The paper introduces a machine learning pipeline using a multi-label segmentation deep network with dynamics convolution to diagnose EoE by identifying and quantifying Eos in whole slide images (WSIs) of esophageal biopsies. The model's significance lies in its ability to accurately segment intact and not-intact Eos, offering a promising automated solution for EoE diagnosis~\cite{daniel2022deep}.

\textbf{Expert-level diagnosis of nasal polyps using deep learning on whole-slide imaging: }The paper introduces an Artificial Intelligence CRS Evaluation Platform (AICEP) that utilizes deep learning algorithms, specifically Resnet50, Xception, and Inception V3, to accurately diagnose eosinophilic chronic rhinosinusitis with nasal polyps (eCRSwNP) using WSIs. The AICEP offers a rapid and objective approach to assess the ratio of Eos to infiltrating inflammatory cells, potentially improving diagnostic and treatment strategies for affected patients ~\cite{wu2020expert}.

\textbf{DEEP LEARNING TISSUE ANALYSIS DIAGNOSES AND PREDICTS TREATMENT RESPONSE IN \\EOSINOPHILIC ESOPHAGITIS:} The paper presents two deep learning models for tissue analysis in EoE: one to identify Eos in esophageal biopsies and the other to classify esophageal tissue as EoE or normal. These models show promise in diagnosing EoE and predicting treatment response and symptoms at remission, potentially improving histologic assessment of patients and guiding clinical management~\cite{javaid2021deep}.

\textbf{Development and technical validation of an artificial intelligence model for quantitative analysis of histopathologic features of eosinophilic esophagitis:} 
The paper presents the development and technical validation of an artificial intelligence (AI) digital pathology model for quantitative analysis of histopathologic features in EoE. The AI model aims to provide standardized and reproducible evaluations of EoE cases, offering potential benefits in improving diagnostic accuracy and patient management~\cite{archila2022development}.

\textbf{Performance of an Artificial Intelligence Model for Recognition and Quantitation of Histologic Features of Eosinophilic Esophagitis on Biopsy Samples:} 
The paper describes the development and evaluation of an artificial intelligence (AI)-based digital pathology model for recognizing and quantifying histologic features related to EoE. The AI model shows promising performance in accurately identifying and scoring key histopathologic features in the EoE spectrum, demonstrating potential significance in improving diagnostic accuracy and assessment of EoE cases ~\cite{archila2023performance}.

\textbf{Eosinophil Detection with Modified YOLOv3 Model in Large Pathology Image:} The paper introduces a modified YOLOv3 object detection method to automatically count Eos in pathology images for the treatment of eosinophilic gastroenteritis. The proposed approach improves accuracy and efficiency in detecting small and crowded Eos, offering a promising solution to reduce the time-consuming and subjective manual counting by pathologists, thereby potentially improving diagnostic and treatment strategies for eosinophilic gastroenteritis~\cite{wang2021eosinophil}.

\textbf{Eosinophilic esophagitis multi-label feature recognition on whole slide imaging using transfer learning:} The paper proposes a transfer deep-learning approach using pre-trained models for automated identification of multiple histologic features of EoE on WSIs, showing promise in improving diagnostic efficiency and reducing manual burdens~\cite{shi2022eosinophilic}.

\textbf{Object-oriented Domain Adaptation for Cell Detection on Pathology Image: } The paper introduces a novel copy-paste data augmentation strategy to improve small cell detection in pathology images using Faster R-CNN. The proposed method enhances the robustness of the detection model and addresses distribution discrepancies at the instance level, leading to significant performance improvements on a self-made Eos pathological dataset~\cite{pan2022object}.

\textbf{A New Approach for Detecting Eosinophils in the
Gastrointestinal Tract and Diagnosing Eosinophilic
Colitis: } The paper introduces an image processing and machine learning approach to diagnose Eosinophilic Colitis in patients by detecting Eos in microscopic biopsies. The proposed method offers a fast and easy tool for pathologists to identify the disease, addressing the underdiagnosed nature of Eosinophilic Colitis and potentially improving patient outcomes~\cite{alzu2021new}.

\textbf{Harnessing artificial intelligence to infer novel spatial biomarkers for the diagnosis of eosinophilic esophagitis: }The paper introduces an AI platform utilizing semantic segmentation to deduce local and spatial biomarkers from eosinophil distribution and basal zone fractions in esophageal biopsies. The model achieved a notable histological severity classification accuracy, highlighting its potential as a decision support system to enhance diagnosis and treatment prediction for EoE patients~\cite{larey2022harnessing}.

 \begin{figure}[ht]
  \centering
  \includegraphics[width=\textwidth]{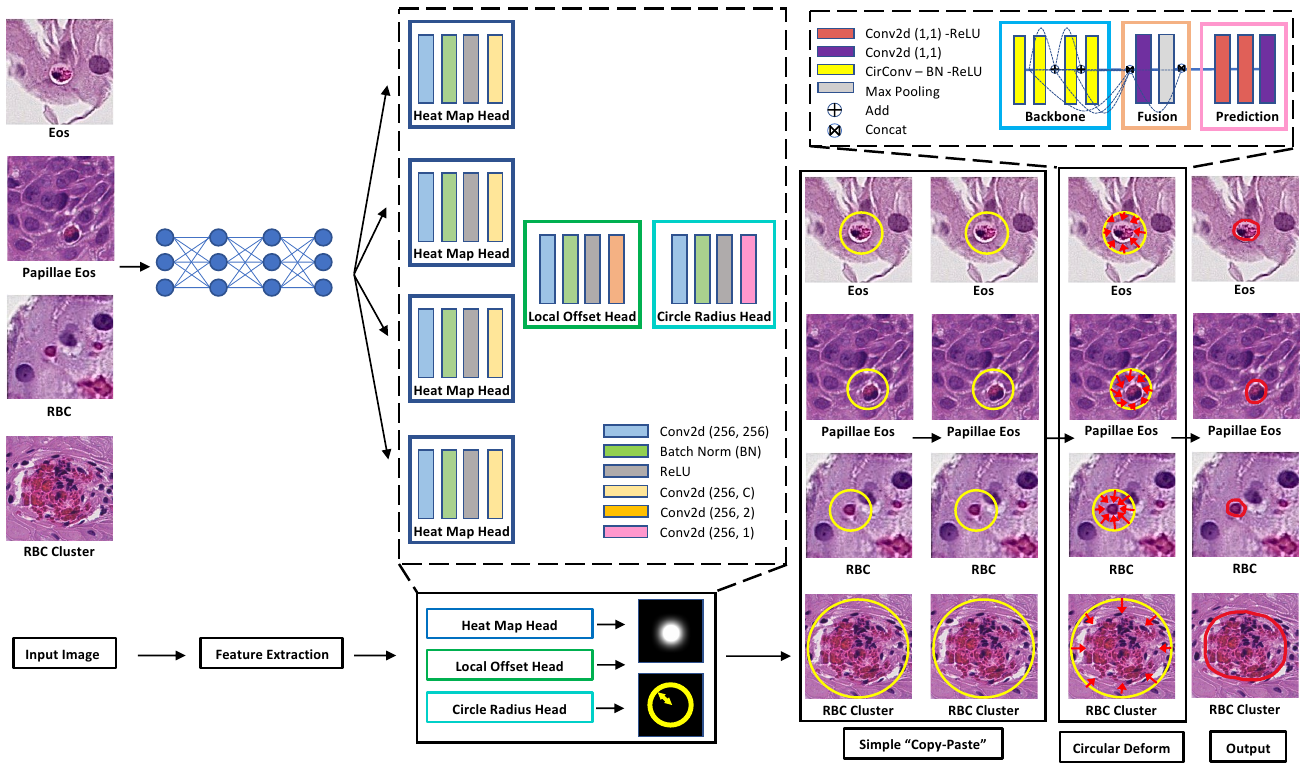}
  \caption{The EoE Data Preparation Pipeline outlines the network structure of the CircleSnake model. By utilizing four heatmap heads, CircleSnake achieves multi-label segmentation. Within the model, CircleNet detects objects and transforms the bounding circle into the final contour through a graph convolutional network (GCN).}
  \label{fig:pipeline}
\end{figure}

\section{METHOD}
\subsection{Phenotype Definition}
In addition to phenotyping Eos, our annotation process includes the identification of Papillae Eosinophils (Papillae Eos) and Red Blood Cells (RBCs) as hard negative cases. Despite their similar appearance, clinical guidance does not consider these phenotypes for use in diagnosis~\cite{liacouras2011eosinophilic,dellon2012diagnosis,muir2021eosinophilic}. By annotating Papillae Eos, RBCs, and RBC Clusters alongside Eos, we enrich the dataset, providing greater data depth. This approach has the potential to yield more accurate and robust Eos instance segmentation results. Specifically, four types of cells were annotated in this study:

\textbf{Eosinophils (Eos)}: Eos are a type of white blood cell that play a crucial role in the immune system's response to allergic reactions and parasitic infections. They are characterized by the presence of large cytoplasmic granules that stain red with eosin, a histological dye. The EoE histological diagnosis threshold is 15 Eos per high-power field~\cite{liacouras2011eosinophilic,dellon2018updated,weidner2009modern,montgomery2006biopsy}.

\textbf{Papillae Eosinophils (Papillae Eos)}: 
Papillae refer to the finger-like extensions of the lamina propria that extend into the epithelium. It is important to note that Eos present within the papillae should not be considered equivalent to intraepithelial Eos and should not be included in the count.

\textbf{Red Blood Cell (RBC)}: 
RBCs are the predominant cell type in the bloodstream, constituting the highest proportion among all cell types. Mature RBCs have a disk diameter of approximately 6.2–8.2 µm, exhibit an oval shape, and possess a biconcave disk structure. They do not have any cell nuclei or organelles, or ribosomes~\cite{nombela2018nucleated,turgeon2005clinical}.

\textbf{Red Blood Cell Cluster (RBC Cluster)}: The label denotes a sizable cluster of RBCs. Due to the difficulty of annotating a cluster of RBCs individually, we made the decision to assign a distinct label to the clusters of RBCs as a new category.

\subsection{Multi-label CircleSnake}
CircleSnake model, is an end-to-end deep learning model, is an enhanced version based on the popular DeepSnake model, specifically designed for accurately detecting ball-shaped medical objects. Unlike the DeepSnake model, the CircleSnake model adopts a circular contour instead of an octagonal contour, resulting in a reduced degree of freedom. This modification enhances the model's capability to effectively solve object detection tasks with improved accuracy and precision~\cite{nguyen2022circlesnake}. In this paper, we extend the CircleSnake from single-label instance segmentation to multi-label instance segmentation.

\subsubsection{Multi-label circle object detection}
The multi-label circle object detection is based on the CircleNet approach~\cite{nguyen2021circle,yang2020circlenet}, known for its high performance and simplicity. Given an input image $I \in R^{W \times H \times 3}$ with dimensions width $W$ and height $H$, the Circle Proposal Localization (CPL) network produces a heatmap $\hat Y \in [0,1]^{\frac{W}{R} \times \frac{H}{R} \times C}$, where $C$ is the number of candidate classes, and $R$ is the downsampling factor. The elements $\hat Y_{xyc}$ in the heatmap represent the probability of the lesion center belonging to candidate class $c$, with $\hat Y_{xyc} = 1$ indicating the center and $\hat Y_{xyc} = 0$ indicating the background. For multi-label object detection, the target center points are represented using a 2D Gaussian kernel to generate the ground truth heatmap $Y$ for each candidate class $c$ (Fig.~\ref{fig:pipeline}):
\begin{equation}
{Y_{xyc} = \exp\left(-\frac{(x-\tilde p_{cx})^2+(y-\tilde p_{cy})^2}{2\sigma_p^2}\right)}
\end{equation}
where $\tilde p_{cx}$ and $\tilde p_{cy}$ are the downsampled ground truth center points for class $c$, and $\sigma_p$ is the kernel standard deviation. The training loss $L_k$ for multi-label detection is given by:
\begin{equation}
    L_k^{(c)} = \frac{-1}{N} \sum_{xy}
    \begin{cases}
        (1 - \hat{Y}_{xyc})^{\alpha} 
        \log(\hat{Y}_{xyc}) & \!\text{if}\ Y_{xyc}=1\\
        \begin{array}{c}
        (1-Y_{xyc})^{\beta} 
        (\hat{Y}_{xyc})^{\alpha}\\
        \log(1-\hat{Y}_{xyc})
        \end{array}
        & \!\text{otherwise}
    \end{cases}
\end{equation}
where $\alpha$ and $\beta$ are hyperparameters, and the summation is taken over all spatial locations and candidate classes.

To propose the $n$ detected center points $\hat{\mathcal{P}} = \{(\hat x_i, \hat y_i)\}_{i = 1}^{n}$, the top $n$ peaks in the heatmap are selected for each candidate class. Each center point $(x_i,y_i)$ contributes to $L_k$, and the offset $(\delta \hat x_i, \delta \hat y_i)$ is computed from $L_{off}$. The bounding circle is formed by combining the center point $\hat{p}$ and radius $\hat{r}$ as follows:
\begin{equation}
\hat{p} = (\hat x_i + \delta \hat x_i ,\ \ \hat y_i + \delta \hat y_i). \quad \hat{r} = \hat R_{\hat x_i,\hat y_i}.
\end{equation}
where $\hat R \in \mathcal{R}^{\frac{W}{R} \times \frac{H}{R} \times 1}$ contains the radius prediction for each pixel, optimized by the loss function:
\begin{equation}
L_{radius} = \frac{1}{N}\sum_{k=1}^{N} \left|\hat R_{p_k} - r_k\right|.
\end{equation}
where $r_k$ is the ground truth radius for each object $k$. The overall objective for multi-label detection is given by:
\begin{equation}
L_{det} = \sum_{c=1}^{C} L_{k}^{(c)} + \lambda_{radius} L_{radius} + \lambda_{off}L_{off},
\label{eq:total_loss}
\end{equation}
where $\lambda_{radius}$ and $\lambda_{off}$ are hyperparameters, and the superscript $(c)$ denotes the loss for each candidate class $c$. In accordance with the findings of Zhou et al.~\cite{zhou2019objects}, we assign the values $\lambda_{radius} = 0.1$ and $\lambda_{off} = 1$.

\subsubsection{Circle contour proposal}
To accommodate multi-label object detection, we extend this methodology based on CircleNet~\cite{yang2020circlenet}. Using bounding circles obtained from the CircleNet approach, we directly derive our new circle contour proposal for each target object (~\ref{fig:pipeline}). By simplifying the contour proposal process and avoiding the complexity of deformation and extreme point-based octagon proposals, our approach ensures simplicity and consistency. The circle proposals are defined by their center point and radius. We uniformly sample $N$ initial points $\{\mathbf{x}^{circle}_i | i = 1, 2,... , N\}$ from the circle contour, starting at the top-most point $x_{1}^{circle}$. Similarly, the ground truth contour is formed by sampling $N$ vertices clockwise along the object boundary (Fig.~\ref{fig:pipeline}). The value of $N$ is set to 128 based on~\cite{peng2020deep}.

\subsubsection{Circular contour deformation}
In the multi-label object detection process of CircleSnake, we analyze a contour with $N$ vertices $\{\mathbf{x}^{circle}_i | i = 1, ..., N\}$ to generate feature vectors for each vertex. These input features $f^{circle}_i$ for a vertex $\mathbf{x}^{circle}_i$ are created by combining learning-based features with the vertex coordinates: $[F(\mathbf{x}^{circle}_i); \mathbf{x}^{circle}_i]$, where $F$ represents the feature maps. These features are considered as a 1-D discrete signal $f: \mathbb{Z} \to \mathbb{R}^D$ along the circular contour. To learn these features, circular convolution, as introduced in~\cite{peng2020deep}, is employed. The circular convolution is mathematically defined as:
\begin{equation}
    (f_N^{circle} \ast k)_i = \sum_{j = -r}^r (f_N^{circle})_{i + j}k_j,
\end{equation}
For the circular convolution operation, we use a learnable kernel function $k: [-r, r] \to \mathbb{R}^D$, and the convolution is denoted by the operator $\ast$, which follows the standard convolution approach. The kernel size is set to nine, as per the methodology in~\cite{peng2020deep}.

To implement the convolution operation, we adopt a graph convolutional network (GCN) approach proposed by Peng et al. in their work~\cite{peng2020deep}. This approach consists of three main components: the backbone, fusion, and prediction. The backbone comprises eight ``CirConv-Bn-ReLU" layers, involving circular convolution with residual skip connections. The fusion block is responsible for combining features from different backbone layers. It achieves this by concatenating the features and passing them through a 1$\times$1 convolutional layer and a max pooling layer to capture information at multiple scales. The prediction head consists of three 1$\times$1 convolutional layers, which produce vertex-wise offsets for the final contour deformation. The loss function for the iterative contour deformation is defined as:
\begin{equation}
L_{iter} = \frac{1}{N}\sum_{i=1}^{N} l_{1}(\tilde x_{i}^{circle} - x_{i}^{gt}).
\end{equation}
where $x_{i}^{gt}$ represents the ground truth boundary point, and $\tilde x_{i}^{circle}$ is the deformed contour point. The contour deformation process is performed iteratively for three iterations, regressing the $N$ offsets according to the approach described in~\cite{peng2020deep}.

\section{DATA AND EXPERIMENTAL DESIGN}

\subsection{Data}

Image patches were cut from 50 WSIs from PediatricEoE dataset with a 40× objective lens. The EoE dataset contains over 12,000 annotations in 50 annotated WSIs. To evaluate the performance of models, the dataset was divided into three subsets: train, val (validation), and test, with a split ratio of 7:1:2, respectively.

The Table.\ref{table1} illustrates the distribution of annotations across these subsets, as well as the total count across all subsets. The ``Train" column represents the number of annotations in the training dataset, with 4,842 annotations for ``Eos", 2,789 annotations for ``Papillas Eos", 426 annotations for ``RBC", and 524 annotations for ``RBC Cluster". Similarly, the ``Val" column displays the counts of annotations in the validation dataset, and the ``Test" column represents the counts in the test dataset. The last row of the table provides the overall totals for each category across all subsets. For example, there are a total of 6,921 annotations for ``Eos", 4,032 annotations for ``Papillas Eos", 566 annotations for ``RBC", and 687 annotations for ``RBC Cluster" in the entire dataset.

\begin{table}[ht]
\caption{Overview of Train, Val, and Test Dataset}
\centering
\begin{tabular}{l@{}c@{\ \ }ccccc}
\toprule
  & Train Set & Val Set & Test Set & All Set \\
\midrule
Eos           & 4842 & 690 & 1389 & 6921 \\
Papillae Eos  & 2789 & 430 & 813 & 4032 \\
RBC           & 426 & 40 & 100 & 566 \\
RBC Cluster   & 524 & 12 & 151 & 687 \\
Total         & 8581 & 1172 & 2453 & 12206  \\
\bottomrule
\label{table1}
\end{tabular}
\end{table}

\begin{figure}[ht]
  \centering
  \includegraphics[width=0.8\textwidth]{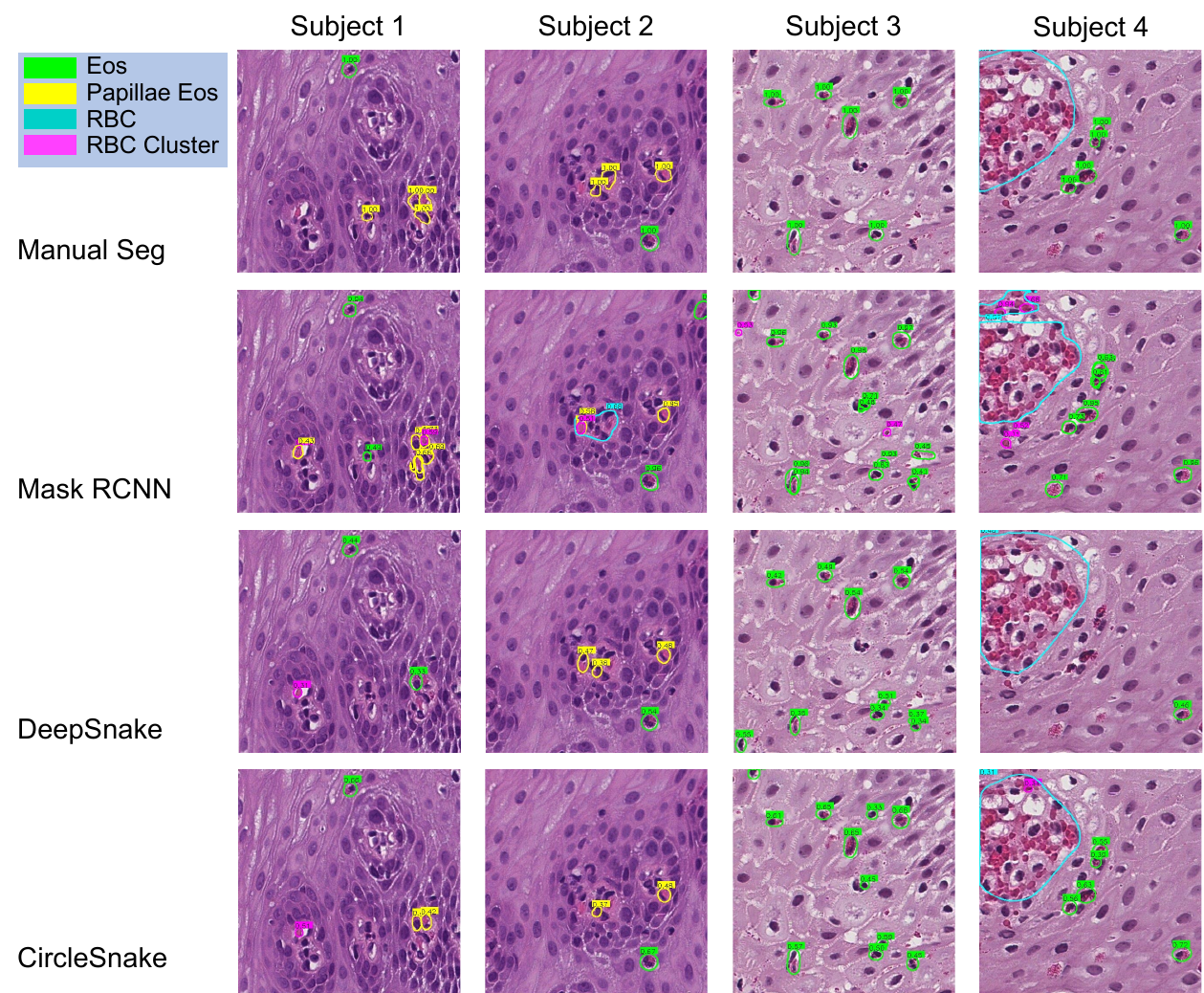}
  \caption{Qualitative segmentation results of different methods (Mask R-CNN, DeepSnake, and CircleSnake), compared to manual annotation}
  \label{fig:output}
\end{figure}

\subsection{Experimental Design}
We sought assistance from a pathologist at Vanderbilt University Medical Center to use QuPath Software for annotating the four different types of cells~\cite{bankhead2017qupath}. In total, we annotated 50 Whole-Slide Images (WSIs) with over 12,000 annotations. Each annotation was classified into one of four categories: Eos, Papillae Eos, RBC, and RBC Cluster. To facilitate further analysis, we utilized QuPath's built-in script editor to convert all the annotations into JSON files. We transformed the JSON files into COCO format so that they could be used as input for our models. We segmented each patch into 512$\times$512 pixels patches and set the downsample to 1 to retain the original resolution. In order to avoid any loss of significant details at the edges, we chose to include an overlap of 256 pixels between adjacent tiles when exporting at the specified resolution. The resulting binary mask was subsequently divided into 512$\times$512 patches that corresponded precisely with the patches in the whole slide images (WSIs). Both the COCO file and the 512$\times$512 patches were utilized as inputs for evaluating the performance of the CircleSnake and Mask R-CNN models.

Utilizing the TileExporter object in QuPath, we retrieved and exported tiles from the original WSIs and the annotated WSIs. Each patch is segmented into 512$\times$512 pixels patches until it covers the whole region. We also set a downsample to define the export resolution. To ensure a smooth transition between tiles and prevent the omission of significant details that might present at the edge of tiles, we used an overlap between adjacent tiles at the export resolution. We produced binary masks for both the ground truth patch and the annotated patch for each patch in our dataset. The generated binary mask is subsequently utilized as input to our models~\cite{bankhead2017qupath}. 

The individual annotations stored in JSON files are converted to COCO format and employed as input for both the CircleSnake and Mask R-CNN models. The validation dataset is utilized to evaluate the performance of different training models and identify the one that achieves the lowest loss, indicating better accuracy and precision. The selected best model from the validation dataset is employed for testing and evaluation on unseen data in the test dataset to assess its generalization and overall performance. For a detailed explanation, please refer to Appendix A.

\begin{table}[ht]
\caption{Results of Mask R-CNN, DeepSnake, and CircleSnake Model\newline}
\centering
\begin{tabular}{l@{}c@{\ \ }ccccccc}
\toprule
  Models &  Backbone & $AP$ & $AP_{(50)}$ & $AP_{(75)}$ & $AP_{(S)}$ & $AP_{(M)}$\\
\midrule
Mask R-CNN & FPN & 0.332 & 0.459 & 0.238 & 0.261 & 0.322 \\
DeepSnake & DLA & 0.222 & 0.331 & 0.121 & 0.133 & 0.418  \\
CircleSnake & DLA & \textbf{0.487} & \textbf{0.705} & \textbf{0.275} & \textbf{0.601} & \textbf{0.638}  \\
\bottomrule
\textbf{\label{table2}}
\end{tabular}
\end{table}

\begin{table}[ht]
\caption{Per-Label Results of Mask R-CNN, DeepSnake, and CircleSnake Model\newline}
\centering
\begin{tabular}{l@{}c@{\ \ }cccccc}
\toprule
  Models &  Backbone & $AP_{(Eos)}$ & $AP_{(Papillae Eos)}$ & $AP_{(RBC)}$ & $AP_{(RBC Cluster)}$ \\
\midrule
Mask R-CNN & FPN & 0.548 & 0.251 & 0.07 & 0.566 \\
DeepSnake & DLA & 0.773 & 0.665 & 0.165 & 0.831 \\
CircleSnake & DLA & \textbf{0.781} & \textbf{0.743} & \textbf{0.363} & \textbf{0.909}  \\
\bottomrule
\end{tabular}
\textbf{\label{table3}}
\end{table}

\section{Results}
Based on the results shown in Table.\ref{table2} and Fig.\ref{fig:output}, the CircleSnake model outperforms both the Mask R-CNN and DeepSnake models in terms of overall performance. It achieves the highest Average Precision (AP) with a value of 0.487, compared to 0.332 and 0.222 for Mask R-CNN and DeepSnake, respectively. CircleSnake also demonstrates superior performance across different IoU thresholds, with higher AP values at IoU thresholds of 0.50 and 0.75.

Moreover, the per-label results provided in Table.\ref{table3} and Fig.\ref{fig:output} reveal that CircleSnake achieves higher AP scores for individual object classes compared to the other models. Specifically, CircleSnake shows higher AP values for Eos, Papillae Eos, RBC, and RBC Clusters, indicating its effectiveness in accurately detecting and segmenting these specific objects.

Overall, these results demonstrate the superiority of the CircleSnake model in accurately identifying and segmenting objects in the given dataset, outperforming both the Mask R-CNN and DeepSnake models. The per-label results further highlight CircleSnake's proficiency in detecting specific cell types and structures within the images.

\section{Conclusion}
In this study, we present an enhanced version of the CircleSnake model that extends its capabilities from single-label to multi-label instance segmentation. By adopting the circle representation, which offers reduced degrees of freedom (DoF=1) compared to traditional polygon representation. Such representation facilitates the downstream deep contour deformation learning process. Our experimental findings demonstrate that the multi-label CircleSnake model outperforms the conventional Mask R-CNN and DeepSnake model in effectively segmenting and diagnosing the four types of cells (Eos, Papillae Eos, RBC, and RBC Cluster). These compelling results highlight the CircleSnake model's potential as a promising approach for accurately identifying EoE phenotypes and potentially finding applications in other instance segmentation scenarios within the realm of digital pathology.

\acknowledgments 
This work has not been submitted for publication or presentation elsewhere. This work is supported in part by NIH R01DK135597(Huo).

\newpage
\appendix
\label{appendix}

\section{Data Preparation Pipeline}

This section will introduce the entire data preparation pipeline for our design in details.

\subsection{Image Annotations Using QuPath}

\subsubsection{Software Installation}
To begin annotating the WSIs, you will need to download the \textbf{QuPath} software. You can download it by following the provided \href{https://qupath.readthedocs.io/en/0.4/docs/intro/installation.html}{link}. Once the software is installed, you can start the annotation process by clicking on the ``Create Project" button. From there, you can open the desired WSIs you want to annotate.

\begin{figure}[H]
\includegraphics[width=0.9\textwidth]{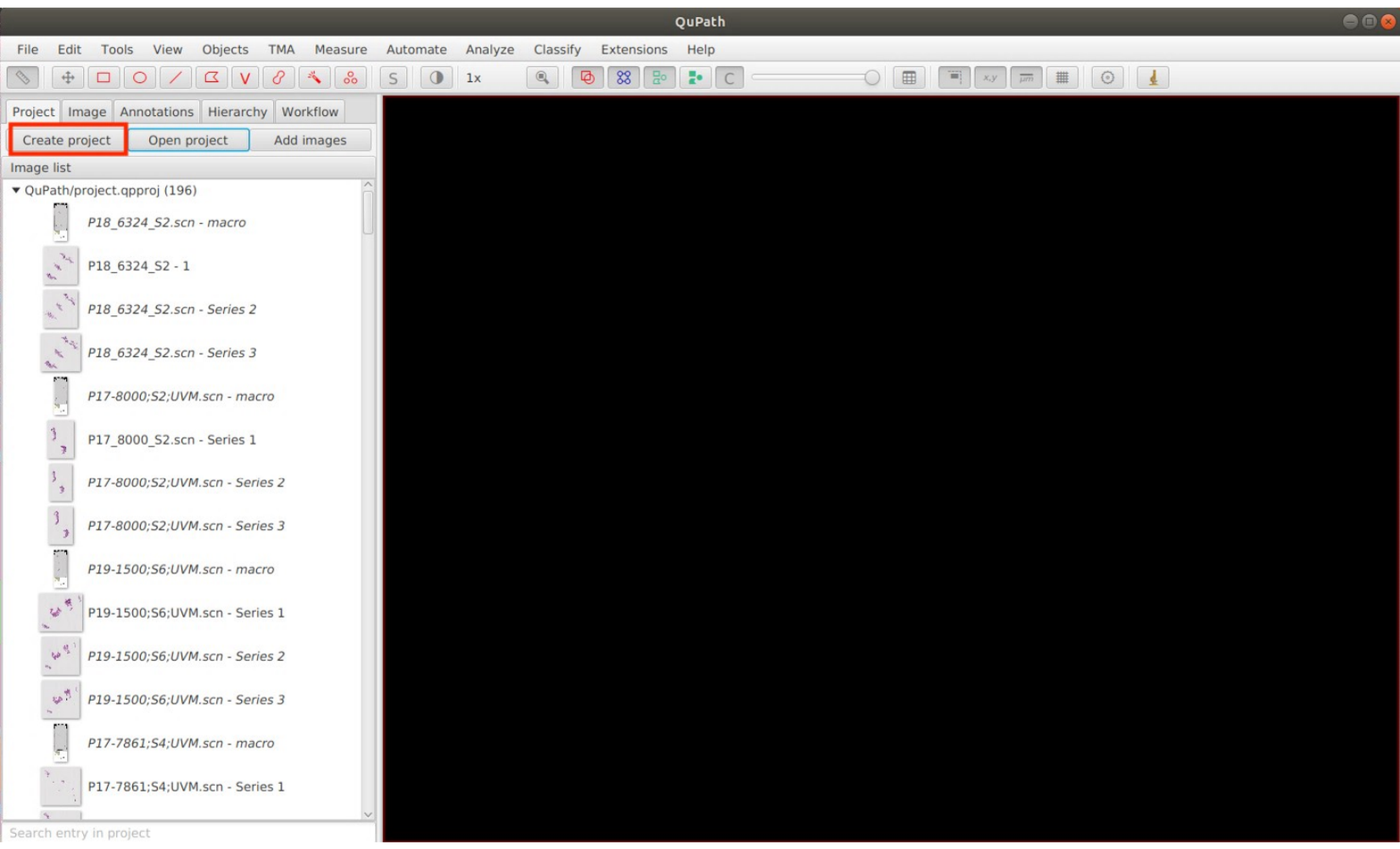}
\centering
\end{figure}

\subsubsection{Open Pre-existing Annotations}
If we have pre-existing annotation file (in .qpdata format) and we want to import the annotation file to our WSIs, we can first click the \textbf{File/save} button and then go to the project directory. In the project directory where your QuPath project is located, you can see an empty annotation file (in .qpdata format). We can replace the empty annotation file by the pre-existing file and reopen the WSIs. Then the annotations will be correctly imported into our project.

\begin{figure}[H]
\includegraphics[width=0.9\textwidth]{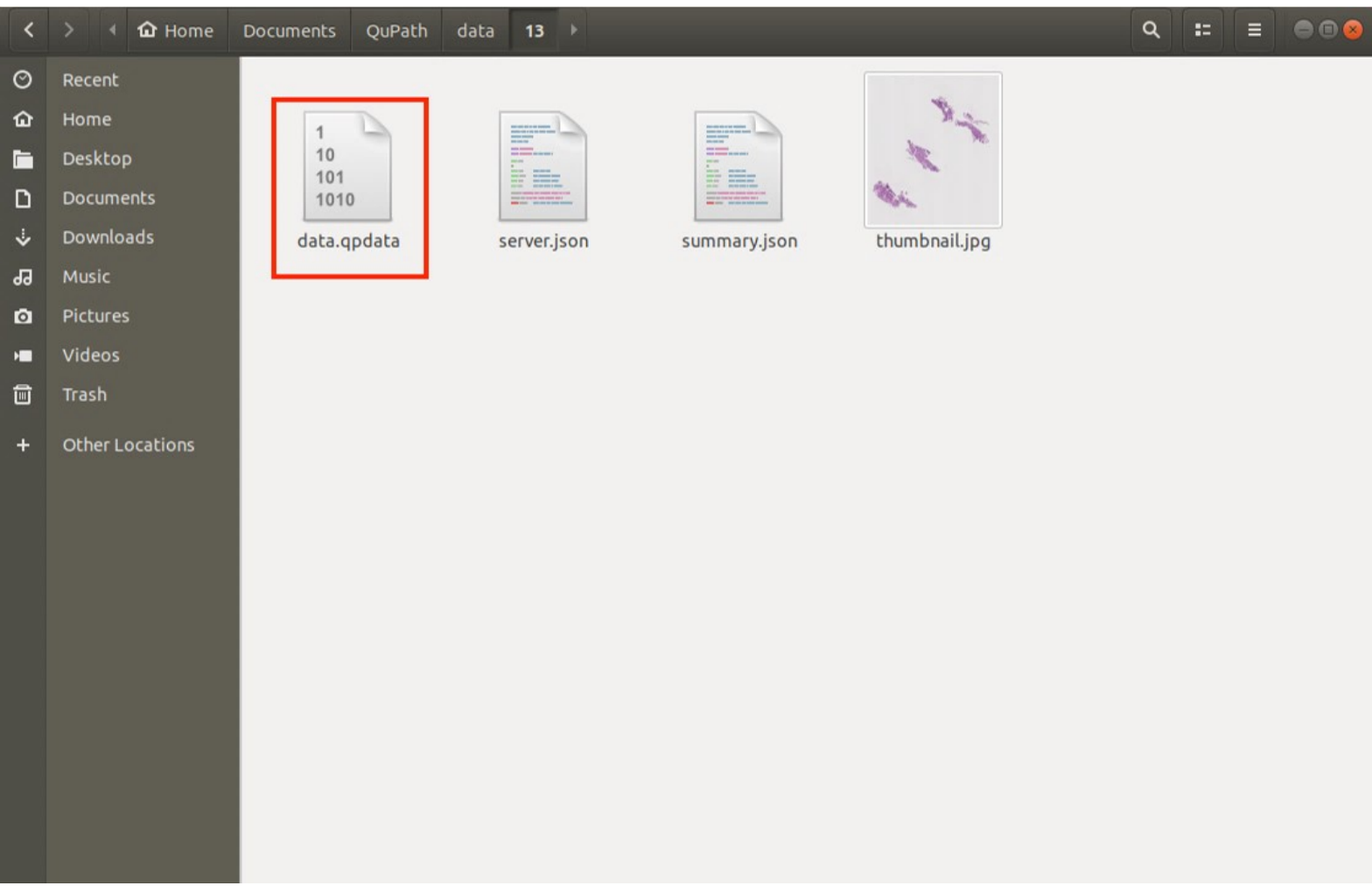}
\centering
\end{figure}

\subsubsection{Annotate from Scratch}
To annotate, we can use the toolbar on the top bar. There are eight different shapes of brush stroke we can use depending on the shape of the objects we want to annotate. Then we can click the shape we want and draw the annotated objects in the WSIs.

\begin{figure}[H]
\includegraphics[width=0.9\textwidth]{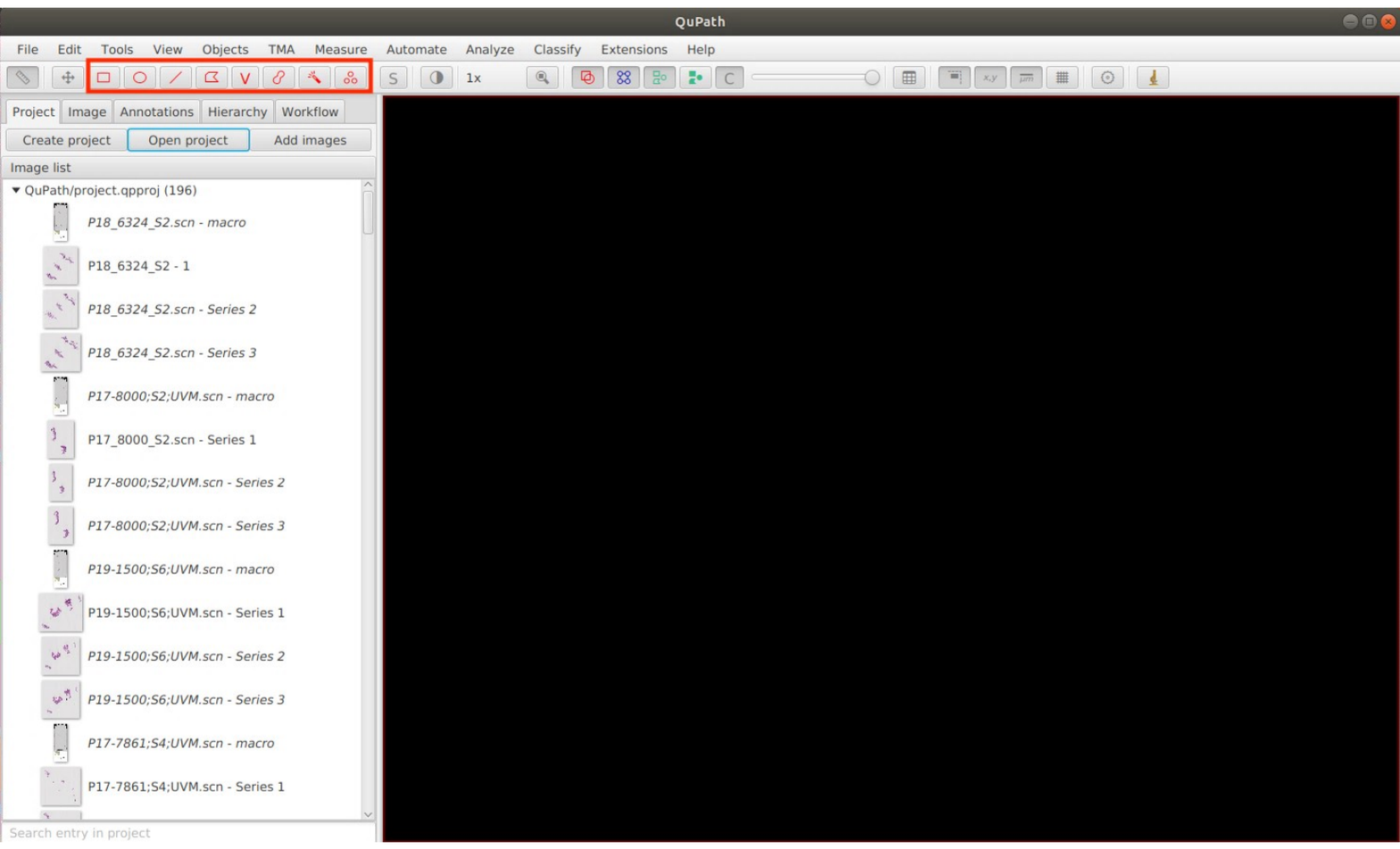}
\centering
\end{figure}

\subsubsection{Multi-label Annotations and Definition}
To access your annotations, navigate to the Annotations button located on the top bar. By clicking on an annotation, you can view the corresponding region and make edits if necessary. To assign a classification to your annotation, select the desired annotations and choose the desired class. Finally, click the \textbf{Set class} button to apply the assigned class.

\begin{figure}[H]
\includegraphics[width=0.5\textwidth]{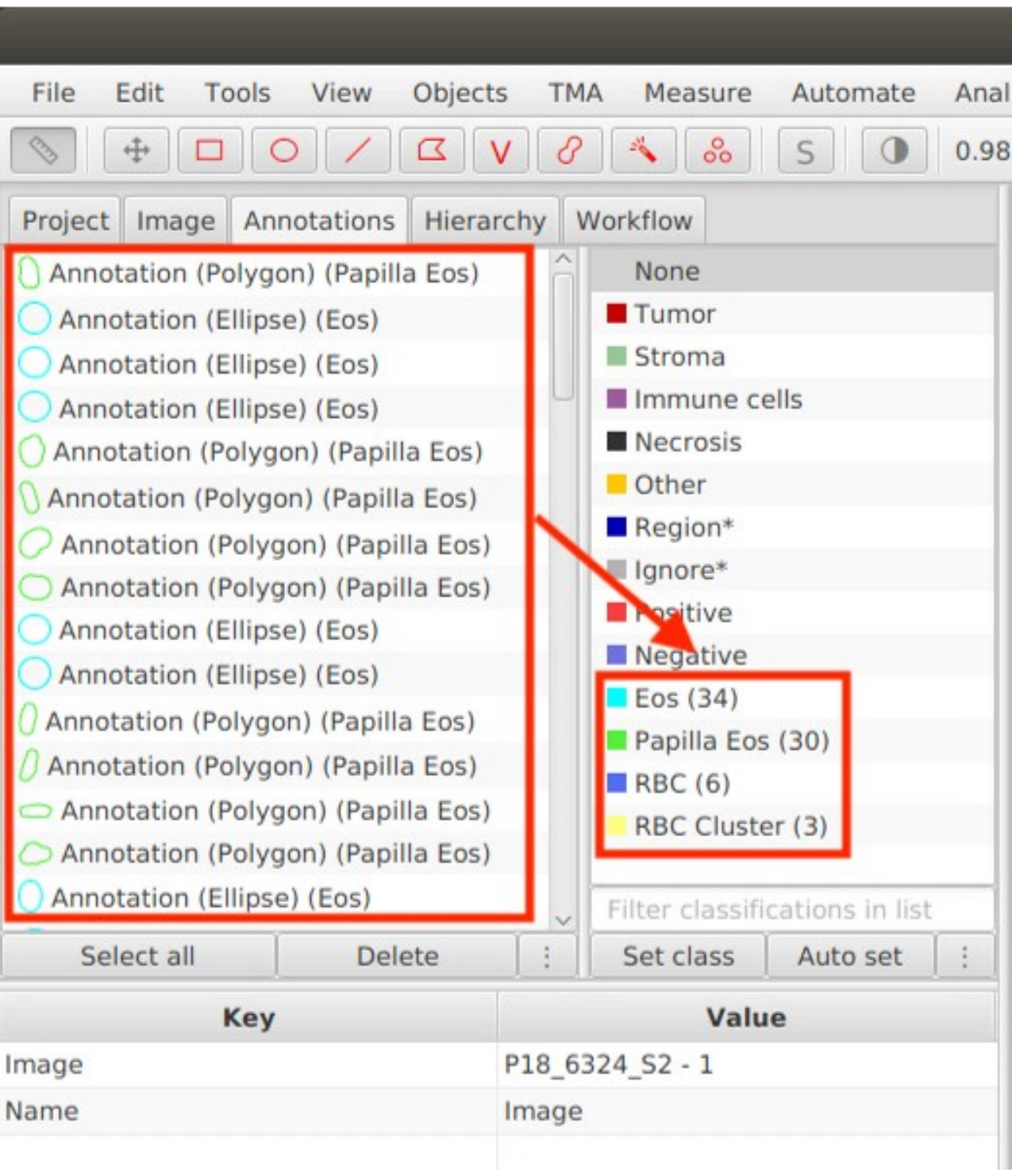}
\centering
\end{figure}

\subsubsection{Convert Annotations into JSON File}
Run the \href{https://github.com/yilinliu610730/EoE/blob/main/src/export_json.groovy}{script} inside QuPath (Automate bar → Show script editor → paste script → run). Make sure to change the \textbf{path} to your directory.

You can also refer to QuPath document for additional help using this \href{https://qupath.readthedocs.io/en/0.4/docs/advanced/exporting_annotations.html#shapes}{link}.

\subsection{Image Segmentation}
\subsubsection{Generate Segmentation}
Run the \href{https://github.com/yilinliu610730/EoE/blob/main/src/export_patches.groovy}{script} inside QuPath (Automate bar → Show script editor → paste script → run). Please ensure that you modify the \textbf{path} in the code to match your directory. Additionally, you have the flexibility to adjust parameters like downsample, tileSize, and pixel overlap according to your specific design requirements. In our implementation, the patch size is set as 512$\times$512 pixels, which means the original WSIs will be segmented into patches of that size.

You can also refer to QuPath document for additional help using this \href{https://qupath.readthedocs.io/en/0.4/docs/advanced/exporting_annotations.html#binary-labeled-images}{link}.

\subsubsection{Generate Contour}
Run the python \href{https://github.com/yilinliu610730/EoE/blob/main/src/json_to_contour.py}{script} to generate contour corresponding to the 512$\times$512 pixels patches (see following figure). As depicted in the figure, the generated contour will be segmented into patches of 512$\times$512 pixels. Each individual annotation will be painted in its respective location within the patch.
If there are multiple annotations within a patch, all of the contours will be generated and placed after the original patch.

\begin{figure}[H]
\includegraphics[width=0.9\textwidth]{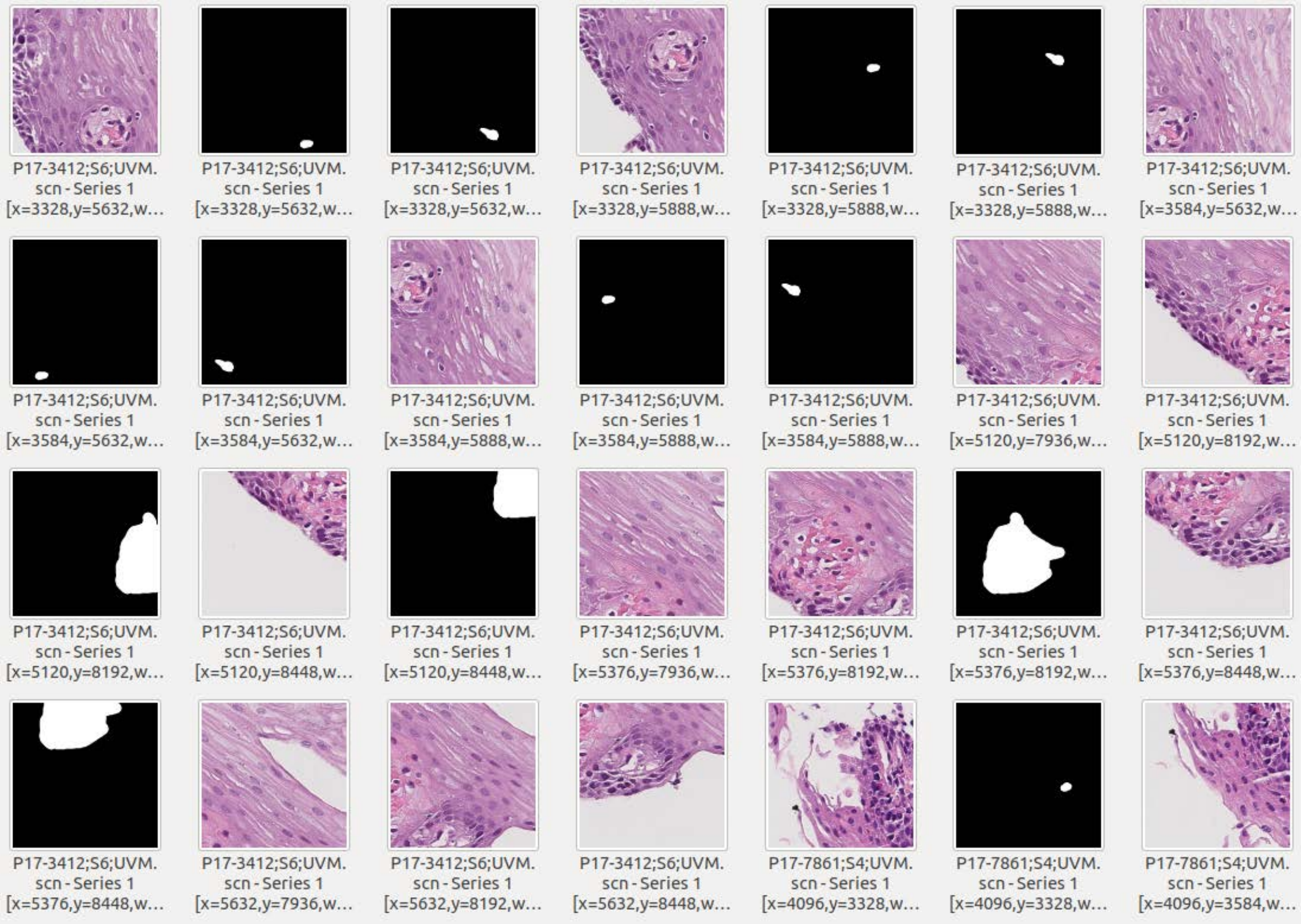}
\centering
\end{figure}

\subsection{Dataset}
\subsubsection{Dataset Splitting}
The EoE dataset consists of more than 12,000 annotations spread across 50 annotated Whole Slide Images (WSIs). In order to assess the performance of models accurately, the dataset was divided into three subsets: train, val (validation), and test. The split ratio used for this division was 7:1:2, respectively. You have the flexibility to adjust the dataset splitting ratio according to your specific design requirements. Feel free to set the splitting ratio that best suits your needs.

\subsubsection{Generate COCO File}
To convert the contour patches we have created into a COCO file, you can run the provided Python script. Please use the following link to access the \href{https://github.com/yilinliu610730/EoE/blob/main/src/class_to_coco.py}{script}. To set the train, val, and test dataset ratios that you have specified, you can update the ``sublist" array as shown below:

\begin{minted}{python}
    sublist["train"] = ["WSI_1", "WSI_2", ...]
    sublist["val"] = ["WSI_3", "WSI_4", ...]
    sublist["test"] = ["WSI_5", "WSI_6", ...]
\end{minted}

Upon running the script, the contour patches will be read and transformed into the COCO file format. Here is an example of how a COCO format file looks:

\begin{minted}[frame=single,
               framesep=3mm,
               linenos=true,
               breaklines=true,
               xleftmargin=21pt,
               tabsize=4]{js}     
{
  "info": {
    "description": "Example COCO file",
    "version": "1.0",
    "year": 2023,
    "contributor": "Your Name",
    "date_created": "2023-06-14"
  },
  "images": [
    {
      "id": 1,
      "file_name": "image1.jpg",
      "width": 512,
      "height": 512
    },
    {
      "id": 2,
      "file_name": "image2.jpg",
      "width": 512,
      "height": 512
    }
  ],
  "annotations": [
    {
      "id": 1,
      "image_id": 1,
      "category_id": 1,
      "segmentation": [[x1, y1, x2, y2, x3, y3, x4, y4]],
      "area": 100,
      "bbox": [x1, y1, width, height]
    },
    {
      "id": 2,
      "image_id": 2,
      "category_id": 2,
      "segmentation": [[x1, y1, x2, y2, x3, y3, x4, y4]],
      "area": 200,
      "bbox": [x1, y1, width, height]
    }
  ],
  "categories": [
    {
      "id": 1,
      "name": "Class 1",
      "supercategory": "Class"
    },
    {
      "id": 2,
      "name": "Class 2",
      "supercategory": "Class"
    }
  ]
}
\end{minted}
\label{json-example}

\subsection{Models}
We utilized two models in our project, namely the CircleSnake model and the Mask R-CNN model. For detailed instructions on how to install these models, you can refer to the following links:

\begin{itemize}
  \item CircleSnake Installation: \href{https://github.com/hrlblab/CircleSnake/blob/main/docs/INSTALL.md}{CircleSnake Installation Guide}
  \item Mask R-CNN Installation: \href{https://detectron2.readthedocs.io/en/latest/tutorials/install.html}{Mask R-CNN Installation Guide}
\end{itemize}

Once you have completed the installation, you can proceed with the training and testing processes by following the specific instructions provided for each model:

\begin{itemize}
  \item CircleSnake Training and Testing: \href{https://github.com/hrlblab/CircleSnake/blob/main/docs/PROJECT_STRUCTURE.md}{CircleSnake Training and Testing Guide}
  \item Mask R-CNN Training and Testing: \href{https://detectron2.readthedocs.io/en/latest/tutorials/getting_started.html}{Mask R-CNN Training and Testing Guide}
\end{itemize}

\subsubsection{Configure Your Model}
For CircleSnake model, to customize the configuration file based on your design, follow these steps:

\begin{itemize}
  \item \textbf{Locate the configuration file}: The config file is located inside the ``CircleSnake/configs" folder. An example of a configuration file is as follows.

  \item \textbf{Understand the structure}: Take some time to familiarize yourself with the structure of the configuration file.

  \item \textbf{Modify the parameters}: Within the configuration file, you'll find different sections and parameters that can be customized. Modify the values of the parameters according to your design requirements.

  \item \textbf{Test the changes}: Launch the CircleSnake application using the modified configuration file to test your changes. Observe the behavior of the game and verify that the customized parameters have taken effect as intended.

  \item \textbf{Fine-tune the parameters:} If needed, repeat the process of modifying the configuration file, testing the changes, fine-tune the parameters according to your design.
\end{itemize}

\begin{minted}[frame=single,
               framesep=3mm,
               linenos=true,
               breaklines=true,
               xleftmargin=21pt,
               tabsize=4]{yaml}     

model: 'coco'
network: 'ro_34'     # Syntax: arch_numOfLayers
task: 'circle_snake' # Determines which network to call
resume: false
gpus: (0,) # Must be a tuple

train:
    optim: 'adam'
    lr: 2.5e-4
    milestones: (60, 80, 100, 150)
    gamma: 0.5
    batch_size: 1
    dataset: 'eoeTrain' # Change this to your dataset
    num_workers: 1
    epoch: 200
    weight_decay: 0.0
test:
    dataset: 'eoeTest' # Change this to your dataset
    batch_size: 1

heads: {'ct_hm': 4, 'radius': 1, 'reg': 2}
segm_or_bbox: 'segm'
ct_score: 0.05
save_ep: 5
eval_ep: 5
               
\end{minted}
\label{configuration file}

\subsubsection{Register Your Dataset}
To register your dataset in the CircleSnake application, you should add the dataset directory and name to the "/CircleSnake/lib/datasets/dataset\_catalog.py" file. Here's an example of how it can be done:

\begin{minted}[frame=single,
               framesep=3mm,
               linenos=true,
               breaklines=true,
               xleftmargin=21pt,
               tabsize=4]{python}     
dataset_attrs = {
    "eoeTrain": {
        "id": "coco",
        "data_root": YOUR_TRAIN_ROOT,
        "ann_file": YOUR_TRAINING_COCO_FILE,
        "split": "train",
    },
    "eoeVal": {
        "id": "coco",
        "data_root": YOUR_VAL_ROOT,
        "ann_file": YOUR_VALITION_COCO_FILE,
        "split": "test",
    },
    "eoeTest": {
        "id": "coco_test",
        "data_root": YOUR_TEST_ROOT,
        "ann_file": YOUR_TESTING_COCO_FILE,
        "split": "test",
    }
}
\end{minted}
\label{Register your dataset}

\subsubsection{Configure Training and Testing Settings}

To train the CircleSnake model, you can create a Run/Debug configuration with the following settings:

\begin{minted}[frame=single,
               framesep=3mm,
               linenos=true,
               breaklines=true,
               xleftmargin=21pt,
               tabsize=4]{python}
               
Script path: 
/CircleSnake/run.py

Parameters:
--cfg_file configs/coco_circlesnake_eoe.yaml
model CircleNet_eoe
train.dataset eosTrain
test.dataset eosTest
pretrain ctdet_coco_dla_2x_converted
debug_train False
train.batch_size 16
               
\end{minted}
\label{Run Training}

To test the CircleSnake model, you can create a Run/Debug configuration with the following settings:
\begin{minted}[frame=single,
               framesep=3mm,
               linenos=true,
               breaklines=true,
               xleftmargin=21pt,
               tabsize=4]{python}

Script path: 
/CircleSnake/run.py

Parameters:
--type evaluate
--cfg_file configs/coco_circlesnake_eoe.yaml
model CircleNet_eoe
test.dataset eosTest
test.epoch 49
ct_score 0.2
segm_or_bbox segm
dice True
debug_test True
save_images True
rotate_reproduce False
               
\end{minted}
\label{Run Testing}

These guides will provide you with the necessary steps and instructions to train and test your models effectively.

\bibliography{report} 
\bibliographystyle{spiebib} 

\end{document}